\documentclass[
 preprint,
 superscriptaddress,
 amsmath,amssymb,
 aps,physrev,
]{revtex4-2}

\usepackage{graphicx}
\usepackage{dcolumn}
\usepackage{bm}

\begin{document}

\title{Push-Pull acousto-optic modulator based on non-suspended thin-film lithium niobate on silicon substrate}

\author{Haorui Ni}
\affiliation{Elmore Family School of Electrical and Computer Engineering, Purdue University, West Lafayette, Indiana 47907, United States}

\author{Sunil Bhave}
\affiliation{Elmore Family School of Electrical and Computer Engineering, Purdue University, West Lafayette, Indiana 47907, United States}

\author{Mengyue Xu}
\email[Contact author: ]{mengyue.xu01@gmail.com}
\affiliation{Elmore Family School of Electrical and Computer Engineering, Purdue University, West Lafayette, Indiana 47907, United States}
\affiliation{Department of Electrical and Computer Engineering, University of Michigan, Ann Arbor, Michigan 48109, USA}

\date{\today}

\begin{abstract}
Acousto-optic modulators (AOMs) are particularly attractive for microwave-to-optical conversion, quantum transduction, and optical frequency manipulation. For these applications, chip-scale AOMs that combine high efficiency, broad bandwidth, and low optical loss are highly desirable. Although suspended and resonant AOMs can enhance modulation efficiency, they typically suffer from stability concerns and limited bandwidth. Here, we demonstrate a non-suspended built-in push-pull AOM on a thin-film lithium niobate (TFLN) on silicon substrate that simultaneously offers high efficiency and relatively broad bandwidth. We further investigate the orientation dependence of electromechanical coupling in X-cut TFLN by fabricating devices with different acoustic propagation directions and identify an optimized orientation for enhanced acousto-optic transduction. Our low-loss device achieves a half-wave voltage-length product of $1.004~\mathrm{V\cdot cm}$ at $0.842~\mathrm{GHz}$ with an interaction length of $400~\mu\mathrm{m}$, together with a relatively wide acousto-optic modulation bandwidth of $132.5~\mathrm{MHz}$. These results pave the way for efficient, practical integrated photonic-phononic links.
\end{abstract}

\maketitle

\section{Introduction}

Acousto-optic modulation has attracted growing interest because traveling acoustic waves provide unique control over both the energy and momentum of light through photoelastic modulation and moving refractive index gratings~\cite{CaiAcoustooptical2019}. Unlike conventional electro-optic or thermo-optic approaches, acousto-optic interactions naturally support functionalities beyond amplitude and phase modulation, including optical frequency shifting~\cite{ShaoIntegrated2020}, nonreciprocal photonic operations such as isolation~\cite{ZhangIntegratedwaveguidebased2024,ShiAcoustoOptic2026,TianMagneticfree2021,SohnDirection2019}, and microwave-to-optical or qubit transduction~\cite{HugotApproaching2026,JiangEfficient2020,BlesinBidirectional2024}. These unique capabilities make acousto-optic devices highly attractive for advanced photonic systems in communication systems, signal processing, and emerging quantum technologies.

Bulk acousto-optic devices generally require relatively long interaction lengths and suffer from limited optical-acoustic overlap, which restricts miniaturization and modulation efficiency. By contrast, integrated piezoelectric platforms can electrically excite acoustic waves on chip through interdigital transducers (IDTs)~\cite{WangDesign2015,ni2026finln}, enabling wavelength-scale confinement and much stronger acousto-optic interaction in a compact geometry. A range of material platforms has been investigated for integrated acousto-optic devices, including GaN~\cite{ZhangIntegrated2026}, AlN~\cite{KittlausElectrically2021}, PZT~\cite{ansari2022light}, and lithium niobate (LN). Among these candidates, LN is particularly attractive because it combines a large piezoelectric response, a strong photoelastic effect, and excellent optical properties. Beyond acousto-optics, LN also supports a strong Pockels effect for high-speed broadband electro-optic modulation and rich nonlinear optical functionalities, making it a versatile platform for multifunctional integrated photonics. The thin-film lithium niobate (TFLN) platform further extends these advantages by enabling low-loss etched waveguides with strong optical confinement, thereby improving optical-acoustic mode overlap. In addition, the rapid development of TFLN for high-performance electro-optic devices and wafer-scale photonic integration~\cite{LiHigh2023a} has established a strong foundation for future integrated acousto-optic systems.

However, many previously reported TFLN acousto-optic devices rely on suspended structures~\cite{ShaoMicrowavetooptical2019,SarabalisAcoustooptic2021,HassanienEfficient2021} or high-sound-velocity substrates such as sapphire~\cite{ZhangOnchip2025,SarabalisAcoustooptic2020} to confine acoustic energy and reduce mechanical damping. Such suspended or non-standard substrate platforms significantly increase fabrication complexity and are difficult to integrate with standard TFLN foundry processes. Achieving efficient acousto-optic interaction in a non-suspended TFLN on insulator platform that remains compatible with electro-optic device fabrication therefore remains a challenge and has not been widely explored.

In this work, we demonstrate a highly efficient push-pull AOM based on a non-suspended TFLN on insulator platform. The device combines the push-pull phase modulation of a Mach-Zehnder interferometer (MZI) with the spatial peak-valley distribution of the acoustic wave generated by an interdigital transducer. By placing the two optical waveguides at positions corresponding to opposite acoustic phases, the acoustic strain induces phase shifts with opposite signs in the two interferometer arms, effectively doubling the modulation efficiency and achieving a half-wave voltage-length product as low as $V_{\pi}L = 1~\mathrm{V\cdot cm}$ with acousto-optic modulation bandwidth of 132.5~MHz. The device is implemented on a standard TFLN on insulator stack without suspended structures, providing a fabrication-friendly and scalable platform for integrated acousto-optic devices.

\begin{figure}
\centering
\includegraphics[width=\linewidth]{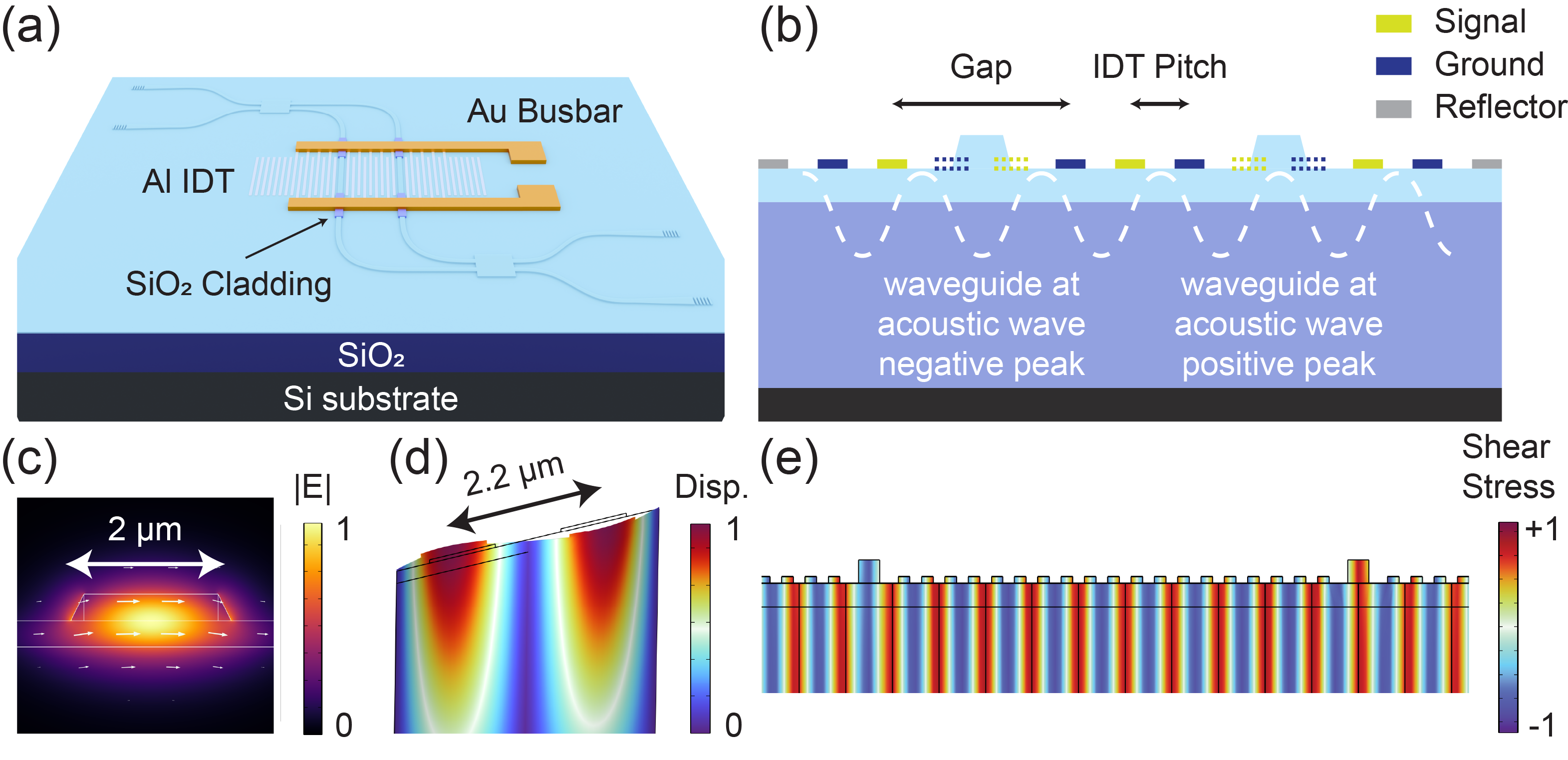}
\caption{\label{fig:1}
(a) Schematic of the proposed push-pull AOM implemented on an X-cut TFLN platform. (b) Cross-sectional view of the acousto-optic modulation region showing the waveguide embedded within the IDT aperture and the omitted IDT fingers near the optical waveguides. (c) Simulated fundamental transverse-electric (TE$_0$) optical mode profile of the rib waveguide at 1550~nm. (d) Simulated displacement field of the shear-horizontal (SH) acoustic wave generated by the IDT obtained from unit-cell finite-element analysis. (e) Full-device acoustic simulation showing the shear strain distribution in the modulation region, illustrating the push-pull acoustic field acting on the two MZI arms.
}
\end{figure}

\section{Device Design}

The schematic diagram of the proposed AOM device is shown in Fig.~\ref{fig:1}(a). The device is fabricated on an X-cut TFLN wafer with 360~nm-thick LN film on a 4.7~$\mu$m buried oxide and silicon substrate. We half-etched LN thin-film using ion milling to form rib waveguides. The AOM consists of an IDT to launch surface acoustic wave (SAW), a pair of metal reflectors to confine the acoustic field, and a balanced MZI to convert the acoustically induced phase modulation into intensity modulation while maintaining a large optical operation bandwidth. The acoustic wave generated by the IDT propagates perpendicularly to the two arms of the MZI. The propagation angle $\alpha$ of the acoustic wave, defined as the angle between the acoustic propagation direction and the crystal Y-axis, is varied for different devices and will be discussed in detail in the following section.

We introduce a novel acousto-optic modulation configuration in which both arms of the MZI are embedded within the IDT aperture to fully exploit the spatial peak-valley distribution of the acoustic field, as illustrated by the cross section of the modulation region in Fig.~\ref{fig:1}(b). In previously reported AOMs, the IDT was placed on only one side of the two MZI arms, requiring the acoustic wave to propagate across the waveguides to establish the desired phase relation~\cite{CaiAcoustooptical2019}. A built-in push-pull configuration was later realized by placing the two waveguides on opposite sides of a central IDT~\cite{WanHighly2022}. However, in these configurations, each waveguide primarily interacts with acoustic waves launched from only one side, and the acoustic field undergoes propagation decay before reaching the optical mode, resulting in weaker acoustic confinement and reduced acousto-optic interaction. In our design, both optical waveguides are located directly inside the IDT aperture, such that each MZI arm is surrounded by IDT fingers on both sides. The waveguides are positioned between the signal and ground electrodes of the IDT, as indicated by the differently colored electrodes in Fig.~\ref{fig:1}(b). This arrangement ensures that the two arms experience acoustic fields with opposite phases corresponding to the peak and valley of the strain field, thereby enabling an intrinsically efficient push-pull modulation configuration. Furthermore, acoustic reflectors are placed outside the IDT region to enhance acoustic confinement and reinforce the mechanical resonance, maximizing the acoustic field interacting with the optical modes.

The optical waveguide and IDT are co-designed to maximize acousto-optic overlap while maintaining negligible metal absorption loss. The waveguide width is chosen to be $2~\mu\mathrm{m}$ to support the fundamental TE$_0$ mode at $1550~\mathrm{nm}$. The corresponding TE optical mode profile is shown in Fig.~\ref{fig:1}(c). The IDT is fabricated using 50~nm-thick aluminum electrodes and is designed to generate an acoustic wavelength of 4.4~$\mu$m, corresponding to an IDT pitch of 2.2~$\mu$m (half of the acoustic wavelength). This configuration excites a shear-horizontal (SH) acoustic wave, as shown in Fig.~\ref{fig:1}(d), which is obtained from unit-cell simulations of the displacement field of the IDT structure. The acoustic half wavelength is designed to be slightly larger than the waveguide width in order to maximize the acousto-optic overlap between the optical mode and the peak acoustic stress field. To minimize optical absorption caused by the metal IDT fingers, the two fingers closest to each waveguide are intentionally omitted from the design. As illustrated in Fig.~\ref{fig:1}(b), the dashed electrodes indicate the omitted fingers, resulting in a gap of 5.5~$\mu$m between the nearest electrodes on both sides of each waveguide while maintaining negligible metal absorption loss (0.2 dB/cm). Figure~\ref{fig:1}(e) shows the full acoustic-wave simulation including the omitted IDT fingers. The results clearly demonstrate the push-pull distribution of the shear stress field acting on the two waveguides.

To electrically connect the IDT fingers, the busbars are routed across the optical waveguides. A SiO$_2$ cladding layer is introduced only in the waveguide-busbar overlap regions to separate the optical mode from the metal busbars and thus minimize excess optical loss (Fig.~\ref{fig:2}). Note that no SiO$_2$ layer is used in the acousto-optic interaction region, in order to avoid acoustic damping and maintain strong acoustic confinement.

\section{Fabrication}

\begin{figure}
\centering
\includegraphics[width=\linewidth]{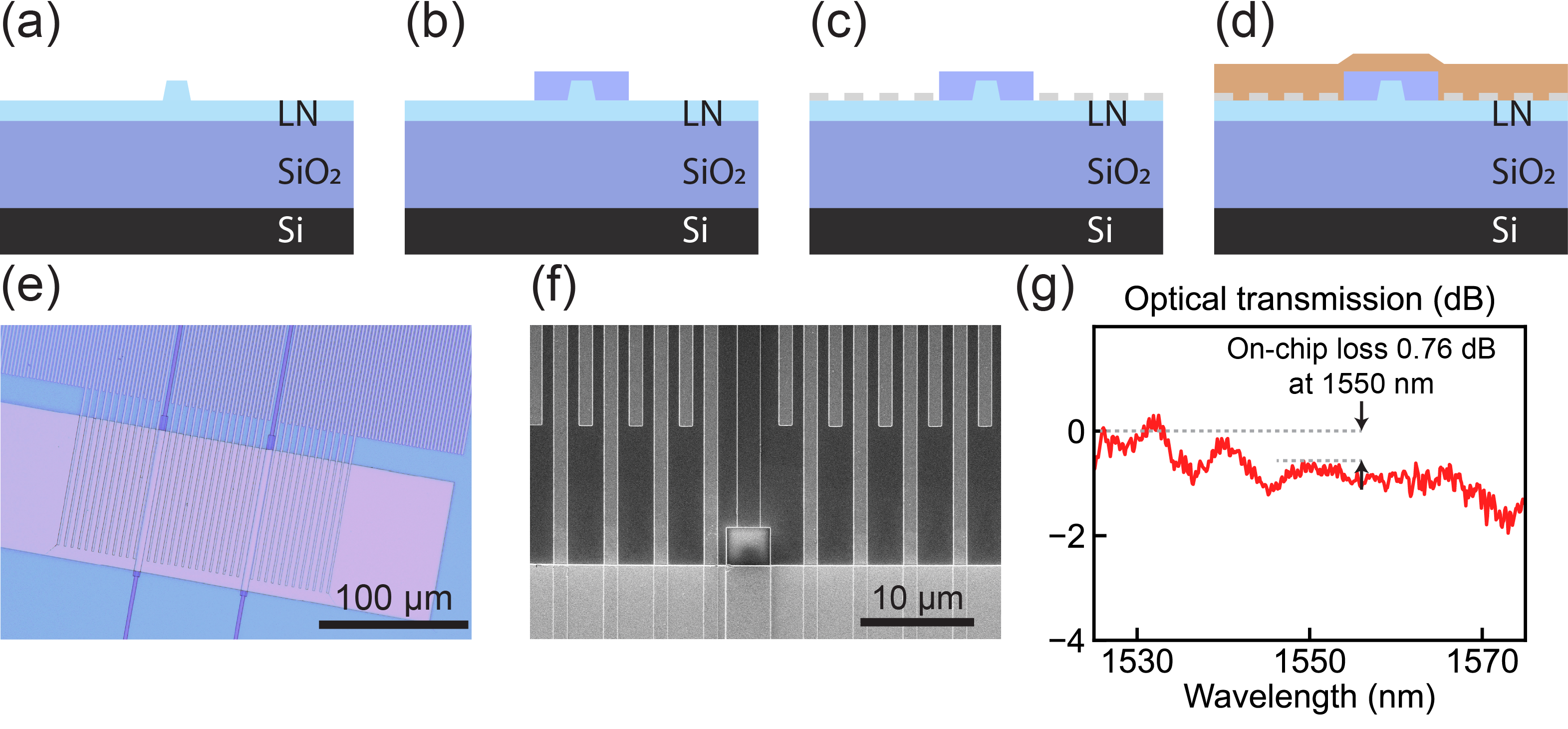}
\caption{\label{fig:2}
Fabrication and optical characterization of the AOM.
(a) Electron-beam lithography (EBL) patterning of the LN rib waveguides using HSQ resist followed by Ar$^{+}$ ion milling.
(b) Deposition and patterning of the HSQ cladding layer.
(c) Definition of the Al IDT electrodes using EBL and lift-off.
(d) Fabrication of the Au busbars using the same lift-off process.
(e) Optical microscope image of the fabricated AOM device.
(f) Zoomed-in scanning electron microscope (SEM) image of the acousto-optic modulation region.
(g) Measured transmission spectrum of the balanced MZI structure, showing an extracted on-chip propagation loss of 0.76~dB at 1550~nm.
}
\end{figure}

The AOM device is fabricated on an X-cut TFLN wafer (NanoLN) consisting of a 360~nm thin-film LN layer on a 4.7~$\mu$m buried oxide and thick silicon substrate. The fabrication process flow is illustrated in Fig.~\ref{fig:2}(a)--(d). The optical waveguides are defined using electron-beam lithography (EBL) with hydrogen silsesquioxane (HSQ) as the resist and subsequently etched by Ar$^{+}$ ion milling. The rib waveguides are partially etched by 180~nm, leaving a 180~nm LN slab layer. After removing the LN redeposition, a 600~nm-thick HSQ layer is spin-coated on the chip and patterned by EBL to form the cladding layer in the waveguide-busbar overlap regions. The chip is then thermally annealed in oxygen at $520^{\circ}$C for 1~hour to convert the HSQ into silicon dioxide and to repair ion-etching-induced crystal damage in the LN layer. The 50~nm-thick Al electrodes of the IDT are defined using EBL with polymethyl methacrylate (PMMA) resist followed by metal evaporation and lift-off. Finally, a 400~nm-thick Au busbar is fabricated using the same EBL and lift-off process. Optical microscope and scanning electron microscope (SEM) images of the fabricated devices are shown in Fig.~\ref{fig:2}(e) and Fig.~\ref{fig:2}(f), respectively. Grating couplers are designed for fiber-to-chip coupling with a measured coupling loss of approximately 4.5~dB per facet. The on-chip propagation loss is extracted from the measured fiber-to-fiber transmission and is estimated to be 0.76~dB at 1550~nm, as shown in Fig.~\ref{fig:2}(g). In addition, the balanced MZI configuration enables the AOM to operate over a broad optical bandwidth.

\section{Device Characterization}

\begin{figure}[tbp]
\centering
\includegraphics[width=\linewidth]{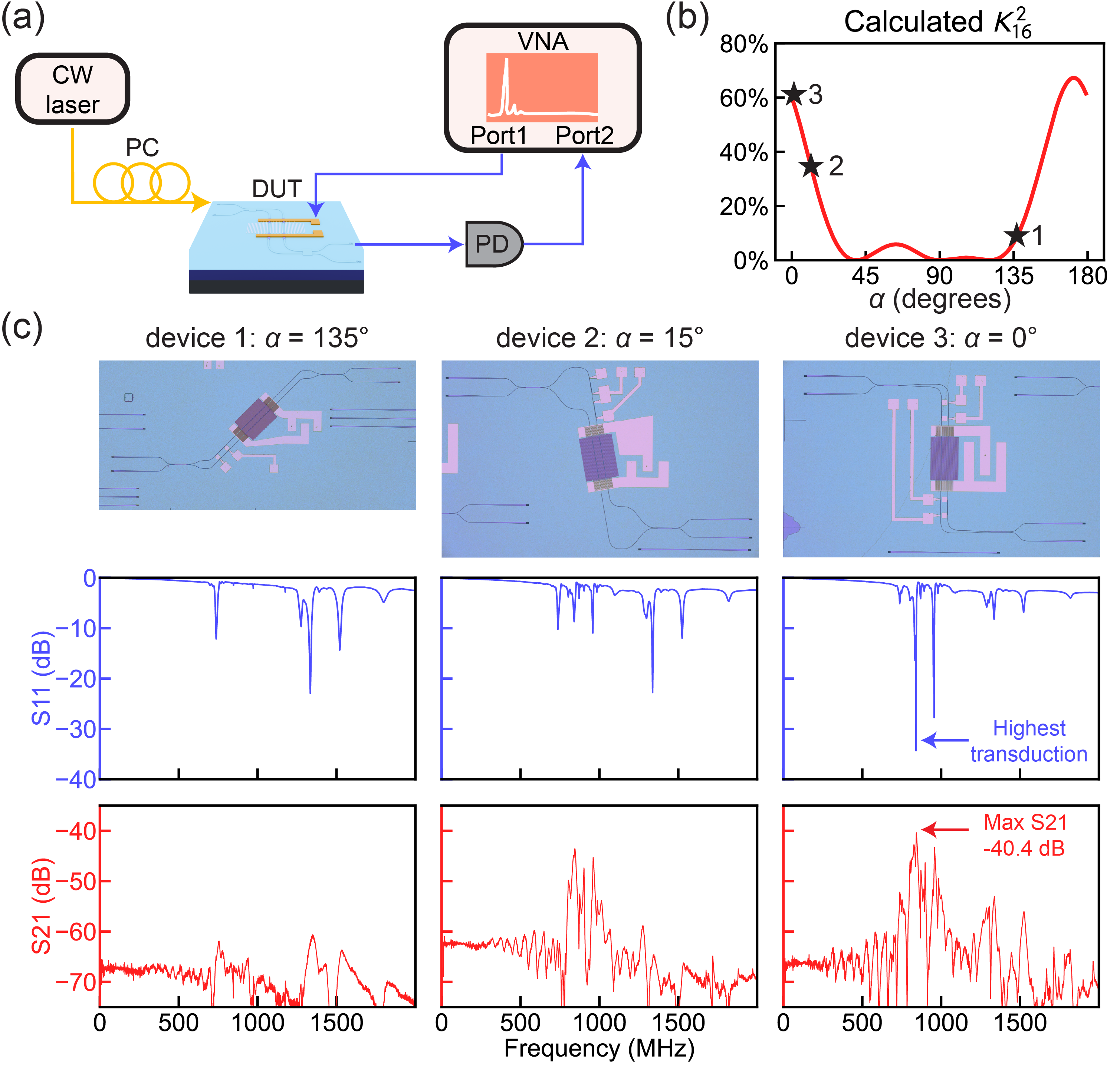}
\caption{\label{fig:3}
(a) Experimental setup for acousto-optic modulation characterization.
(b) Calculated electromechanical coupling coefficient $k_{16}^{2}$ of X-cut lithium niobate as a function of the acoustic propagation angle $\alpha$ with respect to the crystal Y-axis.
(c) Optical microscope images of the fabricated devices with three different acoustic propagation angles: Device~1 ($\alpha=135^\circ$), Device~2 ($\alpha=15^\circ$), and Device~3 ($\alpha=0^\circ$). The corresponding microwave reflection ($S_{11}$) spectra of the IDT and acousto-optic transmission ($S_{21}$) spectra are shown below each device.
}
\end{figure}

Figure~\ref{fig:3}(a) shows the schematic of the experimental setup used to characterize the modulation response of the proposed AOM. Continuous-wave light at 1550~nm is generated by a tunable laser with an output power of 13~dBm. The fiber-to-chip coupling loss is 4~dB per facet for the TE$_0$ mode. An RF signal from a vector network analyzer (VNA, Keysight N5231B) is applied to the IDT through an RF probe to excite the SAW. The acoustically modulated optical signal is detected by a high-speed photodetector and converted into an electrical signal, which is then sent back to the VNA. The microwave-to-acoustic conversion efficiency is evaluated from the microwave reflection coefficient ($S_{11}$) of the fabricated IDT, while the acousto-optic modulation response is characterized by the acousto-optic transmission spectrum ($S_{21}$).

To quantitatively evaluate the acousto-optic coupling under different device orientations, we analyze the material electromechanical coupling coefficient $k_{ij}^2$ of lithium niobate under the quasi-static approximation~\cite{LuRF2021}. The coupling coefficient is defined as
\begin{equation}
k_{ij}^2 = \frac{d_{ij}^2}{\epsilon_{ii}^T s_{jj}^E}
\end{equation}
where $d_{ij}$ is the piezoelectric coefficient, $\epsilon^T$ is the permittivity under constant stress, and $s^E$ is the elastic compliance under constant electric field. The index $i$ represents the electrical field direction and $j$ denotes the stress component. By comparing $k_{ij}^2$, the optimal combination of acoustic mode, electric field direction, and crystal orientation can be identified. Since the proposed device targets shear-horizontal (SH) acoustic waves, the electromechanical coupling coefficient $k_{16}^2$ of X-cut LN is calculated for different acoustic propagation angles $\alpha$, defined as the angle between the acoustic propagation direction and the crystal Y-axis. The calculated $k_{16}^2$ as a function of $\alpha$ is shown in Fig.~\ref{fig:3}(b). Based on this analysis, three orientations with $\alpha = 135^\circ$, $15^\circ$, and $0^\circ$ are selected to fabricate Device~1, Device~2, and Device~3, respectively, as shown in the optical microscope images in Fig.~\ref{fig:3}(c). The optical waveguide orientation is adjusted accordingly to maintain perpendicular propagation between the optical wave and the acoustic wave.

The microwave response of the IDT is characterized by measuring the reflection coefficient ($S_{11}$) using a VNA. As shown in Fig.~\ref{fig:3}(c), Device~3 exhibits the strongest microwave-to-acoustic transduction. The reflection dip occurs at 0.84~GHz with $S_{11} = -34.3$~dB. The fraction of RF power delivered to the IDT can be estimated as $1-|S_{11}|^2$, which corresponds to approximately $99.96\%$ of the input RF power. The acousto-optic modulation is then characterized by measuring the acousto-optic transmission ($S_{21}$). Among the three devices, Device~3 shows the strongest acousto-optic transduction peak of $-40.4$~dB at 0.84~GHz. Device~2 exhibits a lower acousto-optic response, while Device~1 shows the weakest modulation. This trend agrees well with the calculated $k_{16}^2$ in Fig.~\ref{fig:3}(b), confirming that stronger electromechanical coupling leads to enhanced acousto-optic interaction. Among the three orientations investigated, the device with $\alpha = 0^\circ$ exhibits the strongest electromechanical coupling and acousto-optic transduction, indicating that this orientation provides the most efficient configuration for the proposed AOM on X-cut TFLN.

To further quantify the modulation efficiency, we adopt the half-wave voltage-length product ($V_{\pi}L$) as the figure of merit. The value of $V_{\pi}L$ is extracted from the measured $S_{21}$ spectrum using the following equation~\cite{HassanienEfficient2021}:
\begin{equation}
V_{\pi}L = \frac{\pi R_{\mathrm{PD}} P_{0}^{\mathrm{opt}} Z_{0}}{|S_{21}|} L
\end{equation}
where $R_{\mathrm{PD}} = 0.484~\mathrm{A/W}$ is the responsivity of the photodiode at 1550~nm, $P_{0}^{\mathrm{opt}}$ is the output optical DC power, $Z_{0} = 50~\Omega$ is the load impedance of the measurement system, $|S_{21}|$ is the measured AO transduction peak in linear scale, and $L$ is the effective modulation length of the device, which is $400~\mu$m in our design. As shown in Fig.~\ref{fig:3}(c), Device~3 exhibits the highest transduction, corresponding to the lowest $V_{\pi}L$ of $1.004~\mathrm{V\cdot cm}$.

\begin{figure}
\centering
\includegraphics[width=0.5\linewidth]{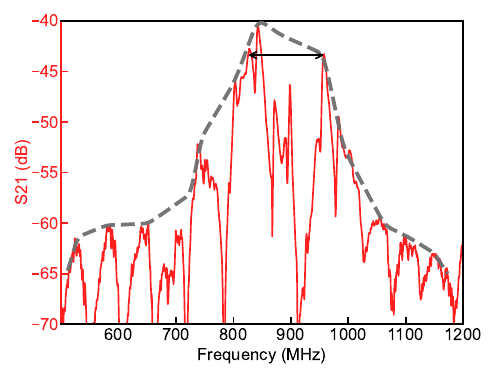}
\caption{\label{fig:4}Zoomed-in acousto-optic transmission ($S_{21}$) spectrum of Device~3 around the main resonance frequency. The dashed curve indicates the envelope used to estimate the modulation bandwidth. The extracted 3-dB acousto-optic modulation bandwidth is 132.5~MHz centered at 0.84~GHz.}
\end{figure}

To further characterize the acousto-optic bandwidth, we zoom in on the acousto-optic transmission ($S_{21}$) spectrum of Device~3 around the resonance frequency, as shown in Fig.~\ref{fig:4}. Here, the 3-dB bandwidth is defined as the frequency interval over which the acousto-optic transduction remains above half of its maximum value~\cite{APLP2025_bw,tadesse2014sub}. The demonstrated 3-dB bandwidth of the AO modulation is 132.5~MHz. The relatively broad modulation bandwidth arises from the excitation of multiple acoustic modes around the center resonance frequency of 0.84~GHz. Because the IDT and the reflectors form an acoustic cavity, several closely spaced acoustic modes participate in the acousto-optic modulation process. Consequently, the modulation response extends over a wider frequency range instead of being limited to a single narrow resonance.

\section{Discussion and Conclusion}

Table~\ref{tab:comparison} summarizes representative piezoelectric acousto-optic intensity modulators~\cite{CaiAcoustooptical2019,KittlausElectrically2021,HuangAcoustoOptic2022,KenningBroadband2025,ansari2022light}. Here, we restrict the comparison to non-suspended, polymer-free devices on Si substrates, which offer mechanical robustness, avoid stability concerns, and remain compatible with standard lithographic processing for scalable fabrication. Even without suspended acoustic cavities or optical resonators, our proposed push-pull MZI delivers the best overall performance in this comparison, exhibiting the lowest measured $V_{\pi}L$ and the largest modulation bandwidth among the listed devices. Specifically, the device achieves a measured $V_{\pi}L$ of $1.004~\mathrm{V\cdot cm}$ at $0.842~\mathrm{GHz}$ and a modulation bandwidth of $132.5~\mathrm{MHz}$, while maintaining a short interaction length of only $400~\mu\mathrm{m}$. This performance is mainly attributed to the enhanced acousto-optic interaction enabled by the push-pull configuration and the stronger acoustic confinement achieved by placing IDT fingers on both sides of each MZI arm, thereby increasing the acoustic field overlap with the optical waveguides.

\begin{table}
\caption{\label{tab:comparison}Comparison of polymer-free, non-resonant, non-suspended piezoelectric acousto-optic intensity-modulated devices on silicon substrates.}
\begin{ruledtabular}
\begin{tabular}{lcccccc}
Ref. & Platform on Si & Structure & $L_{\mathrm{AO}}$ ($\mu$m) & Freq. (GHz) & $V_{\pi}L$ (V$\cdot$cm) & Bandwidth (MHz) \\
\colrule
\cite{CaiAcoustooptical2019} & LN/SiO$_2$ & MZI & 1200 & 0.117 & 5.00 & 0.062 \\
\cite{KittlausElectrically2021} & AlN/Si/SiO$_2$ & Waveguide & 240 & 3.11 & 1.80 & 20 \\
\cite{HuangAcoustoOptic2022} & AlScN/Si/SiO$_2$ & Waveguide & 210 & 3.044 & 1.06 & 3.1 \\
\cite{KenningBroadband2025} & AlN/SiN/SiO$_2$ & Waveguide & 1600 & 0.706 & 1.436 & 0.57 \\
\cite{ansari2022light} & PZT/SiO$_2$ & Waveguide & 180 & 0.576 & 3.35 & 20 \\
This work & LN/SiO$_2$ & MZI & 400 & 0.842 & 1.004 & 132.5 \\
\end{tabular}
\end{ruledtabular}
\end{table}

Future improvements of the proposed non-suspended thin-film LN AOM can further enhance the modulation efficiency and operating frequency. First, integrating optical resonant structures, such as microring or photonic crystal cavities, can significantly enhance the acousto-optic interaction by increasing the effective optical interaction length and optical field confinement. Second, the modulation frequency can be extended to several gigahertz by engineering IDTs with smaller electrode pitch and correspondingly scaled optical waveguide dimensions to maintain strong acousto-optic overlap. Finally, the proposed platform is fully compatible with electro-optic modulators on thin-film lithium niobate. The integration of acousto-optic and electro-optic modulation mechanisms may enable hybrid modulation schemes that combine the advantages of both approaches, providing additional functionality for microwave photonics and integrated signal processing.

In conclusion, we demonstrate a highly efficient push-pull AOM on non-suspended TFLN. By embedding the two arms of an MZI within the aperture of an IDT, the optical waveguides experience acoustic fields with opposite phases enabling efficient push-pull acousto-optic modulation while maintaining compatibility with standard TFLN fabrication processes. We further investigate the orientation dependence of the electromechanical coupling in X-cut TFLN by fabricating and characterizing devices with three different acoustic propagation directions. Both theoretical analysis and experimental results show that the device with $\alpha = 0^\circ$ exhibits the strongest microwave-to-acoustic conversion and acousto-optic transduction among the investigated orientations. The optimized device achieves a $V_{\pi}L$ of $1.004~\mathrm{V\cdot cm}$ at 0.842~GHz with an interaction length of $400~\mu$m and a relatively wide acousto-optic modulation bandwidth of 132.5~MHz. These results demonstrate the feasibility of efficient AOM operation on a non-suspended TFLN-on-insulator platform and provide a practical guideline for orientation engineering in integrated acousto-optic devices.

\begin{acknowledgments}
This material is based upon work supported by the Air Force Office of Scientific Research and the Office of Naval Research under award number FA9550-23-1-0333. Chip fabrication was performed at the Birck Nanotechnology Center at Purdue University. The authors thank the Birck Nanotechnology Center staff for their assistance.
\end{acknowledgments}

\section*{Data Availability}

The data are available from the authors upon reasonable request.

\bibliography{ref}

\end{document}